\documentstyle[aps,preprint]{revtex}
\begin{document}
\draft
\title{$\Sigma$ LIKE TERM IN PION ELECTROPRODUCTION \\ NEAR THRESHOLD}
\author{ Myung Ki Cheoun $^{a)}$
\footnote{Present Adress
: Department of Physics, Yonsei University,
Seoul, 120-749, Korea}
\footnote{e.mail : cheoun@phya.yonsei.ac.kr}, 
Shin Nan Yang $^{a)}$, Bong Soo Han $^{b)}$,
Il-Tong Cheon $^{b)}$}
\address{
a) Department of Physics, National Taiwan University,
Taipei, Taiwan, 10764, R.O.C. \\
b) Department of Physics, Yonsei University,
Seoul, 120-749, Korea}
\maketitle
\begin{abstract}
Sum of the time component of $\Sigma$ term and
the induced pseudo scalar term in axial current is shown to be
the t-channel pion pole in Born terms for pion electroproduction near threshold.
We also show that this $\Sigma$ term represents 
the charged pseudo scalar quark density matrix elements in nucleon and 
manifests itself in the $L_0^+$ amplitude on this reaction.
\end{abstract}
\vspace{1cm}

\section{Introduction}

So far most of the theoretical frameworks 
to describe the pionproduction near threshold by electron or
photon is usually based on Born amplitude \cite{Na91,Dr92}. 
It consists of the contributions 
of all poles to the invariant amplitude, but discarding the
contributions associated with branch cuts, i.e. the resonances
and/or the multi-particle intermediate states \cite{Na91}, thus it 
corresponds to the single particle
intermediate states i.e. s-, u- channels mediated through
nucleon, and t-channel through pion. These diagramms can be calculated by
exploiting the effective lagrangians for the interactions
with a constraint of gauge invariance.
The classical Low Energy Theorem (LET) \cite{Dr92,Ba70}
for pion photoproduction can be obtained this way. Of course one needs to
take the higher order contributions into account
to obtain better theoretical results consistent with the recent
experimental data \cite{Fu96,Berg96} as in the chiral perturbation theory
\cite{Be96,Me96}.

On the other side a current algebra (CA)
approach \cite{Ad68,Am79} is alternative 
to this reaction. It enables us to derive the classical LET with
the help of soft pion limit and the extrapolation to physical pion region. But
the relationship between these two approaches still have a flaw :
one could not derive explicitly the
t-channel pion pole despite of many trials \cite{Dr92,Ra76,Sc91,Do73} 
although the s- and u-channels can be deduced from CA approach. Nevertheless
it does not cause any problem in pion photoproduction because
the t-channel pion pole (we call it t-pole) does not contribute
at the pion threshold.
Consequently the absence of t-pole in CA
does not affect its predictive power at all. However
in case of electroproduction it survives even at the pion threshold 
due to the coulomb polarization of virtual photon and 
contributes to longitudinal amplitudes.

Therefore one needs more careful analyses to apply CA method or 
Born approximation to pion electroproduction. 
Originally in CA approach one has two pole structures related
to pion \cite{Dr92,Sc91}. One of 
them is the induced pseudo-scalar (PS) part in the expectation value 
of axial current (call it as h-pole). Another pole appears
from the time component of $\Sigma$ term ($\sigma$-pole).
This $\Sigma$ term is called as chiral symmetry breaking term
due to its operator structure similar
to the $\sigma$ term in $\pi$-N scattering, and has two
components, time and spatial parts.

The concept of $\Sigma$ term in pionproduction 
is originated by Furlan {\it et al.}
\cite{Fu74} and exploited by Nath and Singh \cite{Na89} to show that its
spatial component can be attributed to the discrepancy between
the old experimental data \cite{Ma86} and the classical LET for
$\pi^0$ production. But
the evaluations of the spatial part of this $\Sigma$ term, proportional to
the commutator $ [ D^a (x) , V_i(0) ] $ with $ D^a (x) =
\int d^3 x \partial^\mu A_\mu (x) $, turned out to be strongly model dependent
as discussed in \cite{Ka89,Be91}. Our focus in this report, however, is not
on its spatial part but on its time part. Here it should be reminded that
both parts are fully independent because this $\Sigma$ term is not
self gauge invariant. Moreover the time part can be calculated 
model independently, and has a pion pole structure
($\sigma$-pole). Thus it affects pion electroproduction.
Of course both $\sigma$- and h-poles in CA approach
are entangled in some way to the pion pole in Born terms 
i.e. t-channel pion pole.

But the relations between the poles in these two different approaches are
not clear yet \cite{Am79}. Of
course such a problem could be senseless if one considers the duality 
\cite{Sa95b,Sa95}, supposed to be working in relatively high energy region, 
by which the summation of
s-channels equates the t-channel summation. However if we remind that Born 
approximation has been successful to explain pionproduction with aid of
effective lagrangians approach one needs to theoretically identify the t-pole.

In section II by rederiving a transition amplitude
for pionproduction using real pion limit 
in order to keep the self consistency the
existence of $\Sigma$ term is confirmed. Secondly it is emphasized that
around threshold the sum of
$\sigma$-pole and h-pole in CA approach equals to the t-channel pion
pole in Born terms. It means that the time component of the $\Sigma$ term 
originated from the explicit chiral symmetry breaking effects 
manifests itself as a part of 
t-pole, thus appears as a part of longitudinal
multipole amplitudes in pion electroproduction.
More detailed calculation shows that each half of t-pole contribution
to $L_0^+$ amplitude is attributed 
to $\sigma$- and h- pole contributions, respectively.
Finally 
simple estimation of this effect using the experimental form factors
is given. In section III, it is shown that
the time component of this $\Sigma$ term 
could give us valuable informations about the charged 
PS quark density distribution
in nucleon, as like $\sigma$ term \cite{Ch90} 
in $\pi - N$ scattering, which shows
scalar quark density distribution in nucleon. In section IV we show how 
large it affects the $L_0^+$ amplitude and discuss 
about the possible methods to
extract $\Sigma$ term from the experimental $L_0^+$ amplitude in 
pion electroproduction. A brief summary is done at the final section.

\section{Derivation of $\Sigma$ like term from CA model}

\subsection{Transition amplitude for electropion production}

We demonstrate our 
transition amplitude for electro- and photo-pionproduction using real pion
limit and derive the reciprocal relations between 
the $\sigma$-pole and h-pole embedded in our transition amplitude 
and the t-pole in Born amplitude. 
We start from the hypothesis of partial conservation of axial
current (PCAC),
\begin{equation}
\partial^{\mu} A_\mu^a = - f_\pi m_{\pi}^2 \phi^a~~  , ~
A_\mu^a = g_A {\bar \Psi} \gamma_\mu \gamma_5 {\tau^a \over 2} \Psi
+ f_\pi \partial_\mu \phi^a~, 
\end{equation}
where $f_{\pi}$ is a pion decay constant and $m_{\pi}^2$ is
the square of pion mass. The $f_{\pi} \partial_{\mu} \phi_{\pi}^{a}$ 
is the axial current from pion field satisfying 
the following Klein Gordon equation with nucleon source function,
\begin{equation}
(\Box + m_{\pi}^2) \phi^a = - i g_{\pi N N} {\bar \Psi} \gamma_5 \tau^a \Psi.
\end{equation}
The transition amplitude ${\cal M}_{\nu}^{a}$ for pionproduction 
${V_{\nu}^{\gamma}}(k) + N(p_{1}){\rightarrow}{\pi}^{a}(q) + N(p_{2})$, 
where $a$ is the cartesian isospin (we distinguish the pion charge 
as Greek letter) and $\nu$ is the polarization index, is written in the 
following way within the Bjorken-Drell convention :
\begin{eqnarray}
{\cal M}_{\nu}^{a} & = & ~\int {d^{4}x} {e^{iqx}} (-q^{2}+{m_{\pi}}^{2})
{\langle}{p_{2}}{\vert}
T[{\phi_{\pi}^{a}}(x)V_{\nu}^{\gamma}(0)]{\vert}p_{1}{\rangle} \\ \nonumber
& = & ( { { q^{2} - m_{\pi}^2 } \over {m_{\pi}^2 f_{\pi}  } } )
{\int} d^{4} x e^{iqx} \langle p_{2} \vert T ( {\partial_{\mu} A_{\mu}^{a}
(x) V_{\nu}^{\gamma} (0) }) \vert p_{1} \rangle \\ \nonumber
& = &  ( { { q^{2} - m_{\pi}^2 } \over { m_{\pi}^2 f_\pi  } } )
{\int} d^{4} x e^{iqx} \langle p_{2} \vert \partial_\mu T ( { A_{\mu}^{a}
(x) V_{\nu}^{\gamma} (0) }) - \delta (x_0) [A_0^a (x) , V_\nu^\gamma (0) ]
\vert p_{1} \rangle ~.
\end{eqnarray}
Let us define $C_\nu^a$ and $T_{\mu \nu}^a$ as follows :
\begin{eqnarray}
C_\nu^a &=
{\int} d^{4} x e^{iqx} \delta (x_0) \langle p_{2} \vert   [ A_{0}^{a}
(x), V_{\nu}^{\gamma} (0) ] \vert p_{1} \rangle \\ \nonumber
T_{\mu \nu}^a &=
i {\int} d^{4} x e^{iqx} \langle p_{2} \vert  T [ A_{\mu}^{a}
(x) V_{\nu}^{\gamma} (0) ] \vert p_{1} \rangle ~. 
\end{eqnarray}
Then,
\begin{eqnarray}
f_\pi {\cal M}_\nu^a &= & { {m_{\pi}^2 - q^2} \over m_{\pi}^2} ( C_\nu^a + q_\mu T_{\mu \nu}
^a ) \\ \nonumber
& = & ( C_\nu^a + q_\mu T_{\mu \nu}
^a ) -  {q^2 \over m_{\pi}^2} ( C_\nu^a + q_\mu T_{\mu \nu}
^a ) \\ \nonumber
& = & {\cal M}_{\nu}^{a}(B) + {\cal M}_{\nu}^{a}(C) ~. 
\end{eqnarray}
Here the 2nd term ${\cal M}_\nu^a$(C) does not contribute if one takes soft
pion limit $q^2 \rightarrow 0$. The classical LET in pion photoproduction
can be obtained from ${\cal M}_\nu^a (B)$ i.e. 
the nucleon pole terms can be obtained from $q^\mu T_{\mu \nu}^a$ using the
$N$ and $N \bar{N} N$ intermediate states \cite{Lu68,Er72} and 
the Kroll-Rudermann term, derived from
the minimal coupling of $\pi N N $ interaction in Born term, 
comes from $C_{\nu}^a$. With Goldberg-Treimann (GT) relation
$g_A M = f_\pi g_{\pi N N}$ and 
the equal time commutator (ETC) of axial charge and vector current one 
obtains the following result :
\begin{equation}
C_{\nu}^{a}  = 
{\bar u}(p_{2}) {I_{a} \over 2} [ G_{A}(t){\gamma_{\nu}}{\gamma_{5}}
 +{{G_{P}(t)} \over 2M}{{(k - q)}_{\nu}}{\gamma_{5}}] u(p_{1})~,
\end{equation}
where the momentum transfer is given by
$t ={(k-q)}^{2}$ and $I_{a} = i {\epsilon_{a3b}}{\tau_{b}}$. 
The 2nd term, peculiar to CA approach, is 
the induced pseudo-scalar term. This does not contribute to pion
photoproduction, but makes an important contribution to pion electroproduction 
as will be shown later on. 

If we adopt the real pion limit ${\cal M}_\nu^a (C)$ 
gives the two terms :
\begin{eqnarray}
{\cal M}_{\nu}^{a}(C)
& = & - { 1 \over m_{\pi}^2 }  \int {d^{4}x}~{\partial_0} {e^{iqx}}
{\langle}{p_{2}}{\vert}
~ \delta (x_0) [  \partial_\mu A_{\mu}^{a}(x) ,  V_{\nu}^{\gamma}(0) ]
  ~{\vert}p_{1}{\rangle}  \\ \nonumber
& & + \partial_\alpha
  \int d^{4} x e^{iqx} \langle p_{2} \vert T ( \partial^\alpha
f_\pi \phi^a (x) V_{\nu}^{\gamma} (0) ) \vert p_{1} \rangle~.
\end{eqnarray}
The 1st term leads to the following
contribution, which is called as the chiral symmetry breaking term
because it would go to zero in chiral limit,
\begin{equation}
{  {i q^{0}} \over { {m_{\pi}}^{2}  } } \Sigma_{\nu}^a (\gamma^*  \pi^a)
= {  {i q^{0}} \over { {m_{\pi}}^{2}  } }
{\int}{d^{4}x}{e^{iqx}} {\delta}(x_{0}) {\langle}{p_{2}}{\vert}
 [{\partial}_{\mu} {A}_{\mu}^{a}(x), V_{\nu}^{\gamma}(0)] 
 {\vert}p_{1}{\rangle}~.
\end{equation}
Addition of the 2nd term in ${\cal M}_\nu^a (C)$ to $q_\mu T_{\mu \nu}^a$
in ${\cal M}_\nu^a$(B) yields 
\begin{equation}
i q^\mu {\tilde T}_{\mu \nu}^{a} 
 = i q^\mu {\int}{d^{4}x}{e^{iqx}}{\langle}{p_{2}}{\vert}
 T [{\bar A}_{\mu}^{a}(x)V_{\nu}^{\gamma}(0)] {\vert}p_{1}
{\rangle}~,
\end{equation}
where ${\bar A}_{\mu}^a = 
A_{\mu}^{a}(x) - {f_{\pi}}{\partial}_{\mu}{\phi}^{a}(x) $ is the axial 
currents with the pion axial current subtracted. As a result one does not
retain the pion pole structure in $i q^\mu {\tilde T}_{\mu \nu}^{a} $ any more.

Our transition amplitude is summarized as follows :
\begin{equation}
{\cal M}_{\nu}^{a}
 ={1 \over {f_{\pi}} } C_{\nu}^{a}+
 i { {q_{\mu}} \over {f_{\pi}} } {\tilde T}_{\mu \nu}^{a}
 + {  {i q^{0}} \over { {m_{\pi}}^{2} f_{\pi} } } {\Sigma}_{\nu}^a
 (\gamma^* \pi^\alpha)~.
\end{equation}
This amplitude is identical to Weise's result \cite{Sc91}, which 
is derived from Ward-Takahasi identity. But
we used the real pion limit \cite{Do73}.
The classical LET in photopion production can be derived 
from the 1st and 2nd term similarly to the case in soft pion limit. 
The spatial part of the last term, $\Sigma_{\nu = space}^a$ term, 
are said to contribute to $E_0^+$ amplitude \cite{Sc91} 
in photopion production. For example, the calculation of Nath and Singh
\cite{Na89} showed some interesting effects by using a free quark algebra,
but it was strongly criticized by Kamal \cite{Ka89},
and by Bernstein and Holstein \cite{Be91} in the following respects.
The commutator
$ [ D^a (x) , V_i(0) ] $ in $\Sigma_{space}^a$
should be zero if the model used satisfies PCAC.
Without considering the contributions from the anomalous dimensions at the
loop level due to the gluonic contributions, the axial current by
the free quark algebra does not satisfy PCAC.
Even the free quark calculation shows some model dependent results.
The calculation \cite{Sc91}
of $\Sigma_{space}^{a}$ by chiral nucleon
model showed more or less contribution to $E_0^+$ amplitude, 
whose estimation strongly depends on nucleon size, and that
the longitudinal contribution from $\Sigma_{space}^a$ term
does not affect $L_0^+$ amplitude due to the artificial 
gauge invariance constraint.  

\subsection{The time component of $\Sigma$ term}

If one takes the conservation of vector current (CVC),
equal time commutator ${[ Q_a^5, V_\mu (y) ]}_{x_0 = y_0} = i 
\epsilon_{a 3 b} A_\mu^b (y)$, where $ Q_a^5 = \int d^3 x
 A_0^a (x) $, and its model independent derivative form
 $ [ D^a (x) , V_0 (0) ]  = i \epsilon_{a 3 b} D^b (0) $,
the time component of eq.(8) reduces to
the following nucleon expectation value :
\begin{eqnarray}
{ {i q^0 \Sigma_0^a (\gamma^* \pi^a) } \over {m_{\pi}^2 f_\pi} }& = &
{{ i q^0 } \over { m_{\pi}^2 f_\pi} } 
i  \epsilon_{a 3 b} \langle p_2 \vert \partial ^\mu A_\mu^b (0) \vert p_1
\rangle \\ \nonumber
& = & - {q^0 \over f_\pi} {\bar u} (p_2) \gamma_5 { I_a \over 2}
{ G_P (t) \over 2M} u (p_1)~,
\end{eqnarray}
where we used eqs.(1) and (2) for PCAC and pion source function. The
isospin structure $I_a = i \epsilon_{a 3 b} \tau_b$ means no contribution
of this term to $\pi^0$ production likewise to $C_a^{\nu}$ term in eq.(6).
Here the dependence on the nucleon model is
replaced with the form factor of $G_P (t)$. It means that
one does not need to consider nucleon model if one
uses the experimental form factors of $G_A (t)$
\cite{Dr92} and assumes the pion pole dominance for $G_p (t)$
\cite{Am79,Ch93},
\begin{equation}
G_P (t) = { {4 M^2} \over { m_{\pi}^2 - t}} G_A (t)~,~
G_A (t) = g_A / {(1 - t / \Lambda^2 )}^2~,~ \Lambda \sim 1 GeV ~.
\end{equation}
The momentum transfer at the pion threshold is given as
$ t = {  (k^2 - m_\pi^2  ) / (1 + \mu ) }$ and
$\mu = m_{\pi} / M $. Addition of this term to induced PS part,
the 2nd term in $C_{\nu=0}^a $/$f_\pi$ of eq.(6), gives
\begin{eqnarray}
& & {1 \over f_\pi} {\bar u} (p_2) { I_a \over 2}
  {G_{P}(t) \over 2M} (k_0 - 2 q_0) {\gamma_5} u(p_{1}) \\ \nonumber
& = & {1 \over f_\pi}  {\bar u} (p_2)  I_a 
  { {M G_{A}(t)} \over { ({m_{\pi}^2 - t})  } } (k_0 - 2 q_0) 
{\gamma_5} u(p_{1}) \\ \nonumber
& = & {g_{\pi N N}} {\bar u} (p_2)  I_a 
  { 1 \over {t - {m_{\pi}^2}}  } ( 2 q_0 - k_0) 
{\gamma_5} u(p_{1})~,
\end{eqnarray}
This equals to the time component of
t-pole in Born terms i.e. the t-pole contribution at the pion threshold
can be fully accounted for the sum of $\sigma$-pole and h-pole in CA approach.
Moreover if we calculate transition amplitude by
multiplying photon polarization, the t-pole contribution
at the pion threshold is exactly 
double value of the $\sigma$-pole contribution.
\begin{eqnarray}
& & {g_{\pi N N}}~ {\bar u} (p_2)  I_a 
  { 1 \over {t - {m_{\pi}^2}}  } \epsilon^\mu ( 2 q_{\mu} - k_{\mu}) 
{\gamma_5} u(p_{1}) \\ \nonumber
& = & {g_{\pi N N}}~ {\bar u} (p_2)  I_a 
  { 1 \over {t - {m_{\pi}^2}}  } \epsilon^0 ( 2 q_{0} ) 
{\gamma_5} u(p_{1}) = 2 ~ \epsilon^0 
{  {i q^{0}} \over { {\mu}^{2} f_{\pi} } } {\Sigma}_{0}^a (\gamma^* \pi^\alpha)~.
\end{eqnarray}
It demonstrates that half of the t-pole contribution
at the pion threshold is just the chiral
symmetry breaking effects due to the time component of $\Sigma$ term
in pion eletroproduction, and another half comes from
the induced PS part in $C_{\nu}^a$ term. Of course this can be rewritten
as a pseudo-vector (PV) coupling form if we use the Dirac equation and 
the equivalence relation
between PS and PV coupling constants for $\pi N N$ interaction.
Schaefer and Weise \cite{Sc91} claimed 
that this $\Sigma_0^a$ contribution is equal to
the t-pole in Born terms, but it should be just half of t-pole as shown above.

However it is uncertain \cite{Dr92} 
whether the above relations can be hold generally beyond threshold, because
we need to take the model dependent,
spatial component $\Sigma_{i}^a (\gamma N)$ \cite{Sc91} from commutator
$[ \partial^\mu A_\mu^a (x), V_i^{\gamma} (0)]$ into account as discussed
already.

\subsection{The estimation of $\Sigma_0^a$}

Before discussing the physical meaning of this chiral symmetry breaking 
term, we can extract a value for this term. 
From eq.(11), one can simply get the following result :
\begin{equation}
i \Sigma_0^{\alpha} (\gamma^* \pi^{ - \alpha}) = -
{ m_{\pi}^2 \over { 2 M}} { {G_P (t)} \over {\sqrt{ 2 M ( E_1 + M)  }  }}
<{  { [ \tau_{\alpha}, \tau_3]  } \over 4}    >~ 
\chi_2^+ {\vec \sigma } \cdot {\vec p_1} \chi_1 ~.
\end{equation}
Taking an average value for the spins of initial and final nucleons
\begin{equation}
i {\bar \Sigma}_0^{\alpha} (\gamma^* \pi^{ - \alpha})=
{ m_{\pi}^2 \over {2 \pi M}}
\sqrt{{ { {( W- M)}^2 - k^2}  \over {4 W M} }  } G_P (t)~
<{  { [ \tau_{\alpha}, \tau_3]  } \over 2}    >~ ,
\end{equation}
and eq.(12) for the induced PS form factor and the experimental $G_A (t)$ form
factor, we obtain 38.1 MeV as the value of 
$ \vert i {\bar \Sigma_0^{\alpha}} 
(\gamma^* \pi^{ \pm} ) \vert$ at the
$\gamma$ point i.e. $k^2 = 0$ point. 
The physical implication of this value is discussed at the next section
with explanations for another chiral symmetry breaking terms, the $\sigma$ term
in $\pi - N$ scattering, and the $\Sigma_{space}^a $ term
in $\pi^0$ photoproduction.

\section{ Chiral symmetry breaking }

The chiral symmetry breaking effect
in the $\pi - N$ scattering, so called $\sigma$ term,
gives a well known result \cite{Ch90} about the scalar quark density
distribution of u and d quarks in nucleon,
\begin{eqnarray}
\sigma_{\pi N}^{ab}
& = &i {\int} d^{4} x e^{iqx} \langle p_{2} \vert \delta (x_{0})
[ A_0^a, \partial^{\mu} A_{\mu}^b (x) ] \vert p_{1} \rangle \\ \nonumber
& = &  \langle p_{2} \vert
[ Q^{5a} , [Q^{5b} , H] ] \vert p_{1} \rangle \\ \nonumber
& = & { 1 \over 2} (m_{u} + m_{d}) \delta_{ab}
\langle p_{2} \vert ({\bar u}  u + {\bar d}  d )
\vert p_{1} \rangle~.
\end{eqnarray}
Here the following facts are used : the spatial divergence
of axial current vanishes on integration
over all space, and the axial charges $ Q^{5a} $ are conserved
in the absense of chiral symmetry breaking hamiltonian $H_{int}$.
The total hamiltonian $H = H_0 + H_{int}$,
where $H_0$ is $ {SU(3)}_L \bigotimes {SU(3)}_R $
invariant hamiltonian and
$H_{int}$ is the explicit chiral symmetry breaking term due to quark masses
given by
\begin{equation}
H_{int}
  = m_{u} {\bar u} u
+ m_{d} {\bar d} d + m_{s} {\bar s} s ~.
\end{equation}
If we use the Zweig sum rule from SU(3) symmetry, this $\sigma_{\pi N}$ is
related to the ratio of $\langle N \vert {\bar s} s \vert N \rangle$ and
$\langle N \vert {\bar u} u  + {\bar d} d \vert N \rangle$ \cite{Ch90}. 
Therefore the 
experimental value \cite{Ka88} of $\sigma_{\pi N}$, 
which lies on $ 45 \sim 60$ MeV, gives a valuable insight of 
quark scalar density ${\bar s}
s $ as well as ${\bar u} u + {\bar d} d$ contribution in nucleon.

Using the same method as the
$\sigma$ term in $\pi - N$ scattering we show
that $\Sigma_{space}^{a} (\gamma N) $ can
be expressed as the expectation value of a 
quark tensor current,
\begin{eqnarray}
i \Sigma_{j}^{a} (\gamma N)
& = & i {\int} d^{4} x e^{iqx} \langle p_{2} \vert \delta (x_{0})
[ \partial^{\mu} A_{\mu}^a (x), V_{j}^{\gamma} ] 
\vert p_{1} \rangle \\ \nonumber
& = & {\int} d^{3} x  \langle p_{2} \vert
~[~ [Q^{5a}, H_{int}], V_{j}^{\gamma} ({\vec x}) ] \vert p_{1} \rangle \\ \nonumber
& =  & - { 1 \over 2} (m_{u} + m_{d})
\langle p_{2} \vert \delta_{a 3} ({\bar u} F_{j} u + {\bar d} F_{j} d )
\vert p_{1} \rangle,~with~F_{j} = { 1 \over 2} \epsilon_{jkl} \sigma_{kl}~. 
\end{eqnarray}
This symmetry breaking term, survived in pion photoproduction, should be
gauge invariant by itself because of the manifest gauge invariance of Born 
terms for photoproduction as reported by Schaefer and Weise \cite{Sc91}. 
The redundant term from this condition leads to the cancellation of
the longitudinal part in $\Sigma_{space}^a$ term.

However the time component of sigma term, $\Sigma_0^a$, can be given
fully model independently as the nucleon expectation value
of the axial current divergence (see eq.(11)).
But here instead of using phenomenological
form factors we represent it in terms of quarks using the same techniques
as $\sigma$ term in $\pi - N$ scattering :
\begin{eqnarray}
i \Sigma_{0}^a (\gamma^* \pi^a)
& = & i {\int} d^{4} x e^{iqx} \langle p_{2} \vert \delta (x_{0})
[ \partial^{\mu} A_{\mu}^a (x), V_{0}^{\gamma} ] 
\vert p_{1} \rangle \\ \nonumber
& = & {\int} d^{3} x  \langle p_{2} \vert
~[~ [Q^{5a}, H_{int}], V_{0}^{\gamma} ({\vec x}) ] \vert p_{1} \rangle \\ \nonumber
& = & { 1 \over 2} ( m_u + m_d) \langle p_{2} \vert
 \epsilon_{a 3 c} v_c \vert p_{1} \rangle~,~~with ~~
v_c = -i {\bar q} \lambda_c \gamma^5 q ~.
\end{eqnarray}
This can be rewritten for real pion representation :
\begin{eqnarray}
i \Sigma_0^{\alpha} ( \gamma^* \pi^{- \alpha} )
& = & - { 1 \over 2} (m_u + m_d) 
\langle p_2 \vert {\bar q}_i \gamma_5 {  {[\tau_{\alpha} , \tau_3]
  } \over 2} q_i \vert p_1 \rangle \\ \nonumber
& = &~ {{ (m_{u} + m_{d})} \over \sqrt{2}}
\langle p_{2} \vert - {\bar d} \gamma_{5} u 
\vert p_{1} \rangle ~~ for~~ ( \gamma^* \pi^+) ~ , \\ \nonumber
& or &~  {{ (m_{u} + m_{d}) } \over \sqrt{2}}
\langle p_{2} \vert  ~ {\bar u} \gamma_{5} d 
\vert p_{1} \rangle ~~for ~~ ( \gamma^* \pi^-)~.
\end{eqnarray}
Therefore provided that theoretical or experimental results are known,
one can extract some information for the 
charged PS quark density distribution in nucleon. For example our 
simple estimation for the above terms in section II leads to 38.1 MeV.
The possibility exploiting
experimental results is discussed in next section.

Neutral PS quark density 
$\langle p_2 \vert {\bar u} \gamma_5 u + {\bar d} \gamma_5 d\vert p_1
\rangle $ can be also obtained if the singlet currents $V_\mu^0$ and $A_\mu^0$
in $U(1)_V$ and $U(1)_A$ gauge are taken into account, for instance, in weak
pionproduction. But here we deal with charged PS density
since we consider only the electro-magnetic production of pion on nucleon.

\section{Extraction of $\Sigma$ term from $L_0^+$ amplitude }

As discussed 
already, the contribution of $\Sigma_0^a (\gamma^* N) $ to
transition amplitude for electroproduction is a half of t-pole 
contribution, thus partially contributes to $L_0^+$ amplitude. 
Here we estimate what to extent it affects $L_0^+$
amplitude. We consider the general hadronic transition current matrix element
\cite{Am79,Sch91} for the electroproduction of pion on nucleon. 
They are given by 
\begin{equation}
\epsilon_\mu {\cal M}^{\mu}
= \epsilon_\mu
\sum_{i = 1}^{8} {\bar u}( p_{2}) Q_{i}^{\mu} A_{i} (
\nu, \nu_1, \nu_2)
u(p_{1}),
\end{equation}
where $\nu = (P_i + P_f) \cdot k / 2 M^2, \nu_1 = q \cdot k / 2 M^2, \nu_2
= k^2 / M^2$ and the $u (p_{1}, p_{2})$ are the Dirac spinors
of initial and final nucleons. 
The eight covariant operators $Q_{i}^{\mu}$ are given as follows :
\begin{eqnarray}
Q_{1}^{\mu} &
=& \gamma_{5} \gamma^{\mu},~~~~ Q_{2}^{\mu} = \gamma_{5}
{  {{(P_{i} + P_{f})}^{\mu}}  \over {4 M} }, \\ \nonumber
Q_{3}^{\mu}
& =& \gamma_{5}  { q^{\mu} \over { 2 M}},~~~
Q_{4}^{\mu}  = \gamma_{5}  { k^{\mu} \over { 2 M}},\\ \nonumber
Q_{5}^{\mu}
& =& \gamma_{5}  {   { \gamma^{\mu} \gamma \cdot k -
\gamma \cdot k \gamma^{\mu} }   \over { 4 M }}
,~~~Q_{6}^{\mu}  = \gamma_{5} \gamma \cdot k
{ {  { (P_{i} + P_{f}) }^{\mu} }  \over { 8 M^2} },\\ \nonumber
Q_{7}^{\mu} & =& \gamma_{5} \gamma \cdot k   { q^{\mu} \over { 4 M^2}},~~~
Q_{8}^{\mu} = \gamma_{5} \gamma \cdot k  { k^{\mu} \over { 4 M^2}}~,
\end{eqnarray}
where $k$ and $q$ are the four momenta of initial photon and 
outgoing pion respectively. $P_i$ and $P_f$ are 
the momenta of initial and final nucleons.
The invariant Ball amplitudes $A_i$ are determined from all pole contributions
to transition amplitudes in Born approximation or the chiral effective 
lagrangians in chiral perturbation theory \cite{Be94}. Originally these amplitudes are constrained by
the gauge invariance $k^\mu M_\mu$ = 0, so that one has six independent
invariant amplitudes. Those remained amplitudes can be expressed as CGLN
amplitudes \cite{Ch57}. From them one can directly extract 
the following relation
between the transition amplitudes and the $E_0^+$ and $L_0^+$ 
amplitudes near threshold
using the specialized polarization photon vector \cite{Am79}, 
which has no time component,
\begin{equation}
{ { - e} \over { 4 \pi ( 1 + \mu)} } \epsilon_\mu {\cal M}^{\mu} 
\vert_{thr.} = \chi_f^+ [ E_0^+ {\vec \sigma} \cdot {\vec b} + 
L_0^+ {\vec \sigma} \cdot {\hat k} {\hat k} \cdot {\vec a} ] \chi_i~, 
\end{equation}
where ${\vec a} = {\vec \epsilon} - { \epsilon_0 \over k_0} {\vec k}$ and
${\vec b} = {\vec a} - {\hat k} ( {\hat k} \cdot {\vec a} ) =
{\vec \epsilon} - {\hat k} ( {\hat k} \cdot {\vec \epsilon} )$. 
The result of $L_0^+$ amplitude in terms of $A_i$ amplitude 
is well known
\begin{eqnarray}
E_0^+ \vert_{thr.} & 
= & - { e \over  2M} {[ { M \over { 4 \pi W}} {\sqrt {  {E_i  + M } \over 
{2 M  }}} ~{\cal A}] }_{thr.}~, \\ \nonumber
L_0^+ \vert_{thr.} & 
= & E_0^+ \vert_{thr.} - { e \over  2M} {[ { {E_i - M  } \over {2 M  }}
{ M \over { 4 \pi W}} {\sqrt {  {E_i  + M } \over 
{2 M  }}}~ {\cal B} ] }_{thr.}~,
\end{eqnarray}
where ${\cal A} = A_1 + { \mu \over 2} A_5, {\cal B} = 
- {1 \over 2 } A_2 + A_4 - A_5 + { 1 \over 4} (2 +  \mu) A_6 - { 1 \over 2}
( 2 + \mu) A_8 $. But only $A_3$ and
$A_4$ are survived for t-pole, which are calculated as follows :
\begin{equation}
A_3^{(-)} = - {  {  8 g M^2 F_\pi (k^2) } \over { m_{\pi}^2 - t} }~,~
A_4^{(-)} = {  { 4  g M^2 F_\pi (k^2) } \over { m_{\pi}^2 - t} }~,
\end{equation}
where the coupling constant 
$g_{\pi N N}^{PS}$ is designated as $g$.
One may argue that there is additional term in $A_4^{(-)}$ from the 
gauge invariance of total amplitude. But its contribution is not taken into
account here because we do not need to maintain self 
gauge invariance in t-pole. 
The final value obtained this way is
\begin{equation}
\Delta L_0^+ (\sigma - pole) =
 [ { { e g} \over { 2 M} } { 1 \over { 4 \pi ( 1 + \mu ) }}
\sqrt {  {  { {(2 + \mu)}^2 - \nu_2  } \over { 4 ( 1 + \mu)  }  }}~~ ]
{{ F_\pi ( k^2) } \over \sqrt{2}}
{  {\mu^2 - \nu_2  } \over {\mu^2 ( 2 + \mu) - \nu_2 } }~ for~ \gamma^* \pi^+.
\end{equation}
If we compare to the following 
total $L_0^+$ amplitude \cite{Sch91,Otha}, which was obtained 
from Born approximation,
\begin{eqnarray}
L_0^+ & = &
 [ { { e g} \over { 2 M} } { 1 \over { 4 \pi ( 1 + \mu ) }}
\sqrt {  {  { {(2 + \mu)}^2 - \nu_2  } \over { 4 ( 1 + \mu)  }  }}~~ ]
\cdot \\ \nonumber
& & {\sqrt 2} ~ [ { { F_\pi ( k^2) } }
{  {\mu (\mu + \nu_2)  } \over {\mu^2 ( 2 + \mu) - \nu_2 } }~ 
 - G_E^n { \nu_2 \over { 2 - \nu_2}} - { 1 \over 2} \mu ]
~~for~ \gamma^* \pi^+,
\end{eqnarray}
and let 
\begin{equation}
\alpha = {{ \Delta L_0^+ (\sigma - pole) } \over {L_0^+  }}~,
\end{equation}
then $\alpha_{theory}$ for $\gamma^* \pi^+$ at $\gamma$ point is
about ${ 1 \over 2}$. Of course one can use more refined theory
for $L_0^+$ amplitude to extract $\alpha_{theory}$ \cite{Be96}. 
In order to directly measure this $\sigma$-pole contribution 
, $\Delta L_0^+ (\sigma - pole)$, 
from the experiments one has to separate t-pole contribution from the
whole amplitude. Such a separation is a formidable task in experimental
side. Even the total $L_0^+$ amplitude for $\pi^+$ electroproduction is
not measured yet despite of a try at Saclay \cite{Ch93}. 
Therefore the future experimental results for $L_0^+$ amplitude in $\pi^+$ electroproduction near 
threshold would give semi-empirical value for $\Delta L_0^+ (\sigma - pole)$,
\begin{equation}
\Delta L_0^+ ( \sigma - pole)_{semi-emp.} = L_0^+ (exp.) \cdot 
\alpha_{theory}~.
\end{equation}
If one uses this value at the r.h.s. of eq.(24), one can deduce 
$\Sigma_0^a$ from the l.h.s.. Since the PS quark distribution on
nucleon is closely related to this quantity, 
the $L_0^+ (\sigma - pole)$ amplitude from the future would-be
experments could give invaluable informations about nucleon
structure. Also it could be an important test for our
theoretical results i.e. 
38.1 MeV for $ \vert i {\bar \Sigma_0^{\alpha}} 
(\gamma^* \pi^{ +} ) \vert$ ( see eqs. (16) and (21) ) in section II and III, from experiments.

\section{Conclusion}

The t-channel pole in Born terms contributes significantly to 
eletroproduction. At the pion threshold such a t-pole contribution to
transition amplitude is explained as a sum for those 
of h-pole and $\sigma$-poles in CA approach.
Both contributions are shown to be a just half of t-pole. 
The former (h-pole) comes
from the induced PS part in nucleon expectation value of axial current. The
latter ($\sigma$-pole) is chiral symmetry breaking effect similar to
$\sigma$ term in $\pi - N$ scattering. It means that
chiral symmetry breaking effect explicitly appears as a t-pole contribution
to electroproduction at the pion threshold, thus gives a sizable
contribution to $L_0^+$ amplitude.

From the viewpoint of quark density distribution on nucleon this chiral
symmetry breaking represents a charged PS quark density, which is 
estimated as 38 MeV by exploiting the experimental form factors. 
The future experimental result for $L_0^+$ amplitude is expected to
make some conclusive remarks for this quantity. 
Weak pionproduction can be a good guide about the 
neutral PS quark density distribution on nucleon.

\vskip1cm

{\bf Acknowledgement}

We are grateful to Drs. Su-H Lee, M.Maruyama, K.Saito for their valuable 
comments and fruitful discussions on this work. The work of M.K.Cheoun
were supported by National Science Commeettee of ROC, grant number 
No.85-2811-M002-021 and by Ministry of Education (MOE) of Korea. 
The work of Il-Tong Cheon was supported by the
Basic Science Research Institute Program, MOE of
Korea, No. BSRI-97-2425.


\begin{references}

\bibitem{Na91} A.Nagl, V.Devanathan and H.Ueberall, 
"Nuclear Pion Photoproduction", Springer-Verlag, Berlin, (1991).

\bibitem{Dr92} D.Drechsel and L.Tiator, J.Phys. {\bf G18}, 449(1992).

\bibitem{Ba70} P.de Baenst, Nucl. Phys., {\bf B24}, 633(1970).

\bibitem{Fu96} M.Fuchs {\it et al.}, Phys. Lett., {\bf B368}, 20(1996).

\bibitem{Berg96} J.C. Bergstrom {\it et al.}, Phys. Rev., {\bf C53}, R1052(1996).

\bibitem{Be96} V.Bernard, N.Kaiser and Ulf-G. Meissner, Z. Phys. {\bf C} (to
be published) (1996) and V.Bernard {\it et al.}, Phys. Lett., {\bf B282}, 
448(1992) and Phys. Rev. Lett., {\bf 70}, 387(1993).

\bibitem{Me96} Ulf-G. Meissner, Comm. on Nucl. and Part. Phys.,
Vol.21, No.6, 347(1995).

\bibitem{Ra76} G. M. Raduskii, V. A. Serdyutskii and A. N. Tabachenko,
Sov. J. Nucl. Phys. , {\bf 23}, 2, 212(1976); {\bf 26}, 2, 163(1977);
{\bf 24}, 2, 208(1976).

\bibitem{Ad68} S.L.Adler and R.F.Dashen, Current Algebra and Applications 
to Particle Physics, Benjamine, New York, (1968).

\bibitem{Am79} E.Amaldi, S.Fubini and G.Furlan, Pion-Electroproduction (
Spring Verlag, Berlin, 1979).

\bibitem{Sc91} T.Schaefer and W.Weise, Phys. Lett, {\bf B250}, 6(1990),
and Nucl. Phys., {\bf A534}, 520(1991).

\bibitem{Do73} Norman Dombey and B.J.Read, Nucl. Phys., {\bf B60}, 65(1973).

\bibitem{Ch93} Seonho Choi {\it et al.}, Phys. Rev. Lett, {\bf 71}, 3927(1993).

\bibitem{Fu74} G.Furlan, N.Paver, and C.Verzegnassi, Nuovo Cimento {\bf 20},
295(1974); {\bf 62}, 519(1969); {\bf 70}, 247(1970).

\bibitem{Na89} L.M.Nath and S.K.Singh, Phys. Rev., {\bf 39}, 1207(1989).

\bibitem{Ka89} A.N.Kamal, Phys. Rev. Lett., {\bf 63}, 2346(1989).

\bibitem{Be91} A.M.Bernstein and B.R.Holstein,
Comm. Nucl. Part. Phys., {\bf 20}, 197(1991).

\bibitem{Ma86} E.Mazzucato {\it et al.}, Phys. Rev. Lett., {\bf 57}, 3144(1986).

\bibitem{Sa95b} B.Saghai and F.Tabakin, Phys. Rev., {\bf C53}, 66(1996).

\bibitem{Sa95} J.-C.David, C.Fayard, G.H.Lamot and B.Saghai, 
Phys. Rev., {\bf C53}, 2613(1996).

\bibitem{Ch90} T.P.Cheng and Ling-Fong Li, Gauge Theory in Elementary 
Particle Physics, (1989).

\bibitem{Sch91} S.Scherer and J.H.Koch, Nucl.\ Phys.\ {\bf A534}, 461(1991).

\bibitem{Otha} K.Ohta, Phys. Rev., {\bf C47}, 2344(1993), 
Phys. Rev., {\bf C46}, 2519(1992) and Phys. Rev., {\bf C40}, 1335(1989).

\bibitem{Ch57} G.R.Chew, M.L.Goldberger, F.F.Low and Y.Nambu,
Phys. Rev., {\bf{106}}, 1345(1957).

\bibitem{Ch95} M.K.Cheoun, B.S.Han and Il-Tong Cheon, submitted, 
Jour. Kor. Phys. Soc (JKPS), (1997).

\bibitem{Er88} T.Ericson and W.Weise, Pions and Nuclei, Calendon Press, Oxford(1988).

\bibitem{Lu68} D.Lurie, Particles and Fields, Interscience Publishers, N.Y. London Sydney, (1968).  

\bibitem{Er72} T.Ericson and M.Rho, Phys. Rep., {\bf 5}, No2., 57(1972).
 
\bibitem{Ch89} T.P.Cheng and Ling-Fong Li, Phys. Rev. Lett, 
{\bf 62} 1441(1989).

\bibitem{Ka88} D.B.Kaplan and A.Manohar, Nucl. Phys., {\bf B310}, 527(1988).

\bibitem{Ja90} A.L.Jaffe and A.Manohar, Nucl. Phys., {\bf B337}, 509(1990).


\bibitem{No90b} S.Nozawa and  T.-S.H.Lee, Nucl. Phys. {\bf A513}, 511(1990) and
{\bf A513}, 543(1990).

\bibitem{We92} T.P.Welch {\it et al.}, Phy. Rev. Lett., {\bf 69}, 2761(1992).

\bibitem{Ba61} J.S.Ball, Phys.\ Rev.\ {\bf{124}}, 2014(1961).

\bibitem{Be94} V.Bernard {\it et al.}, Phys. Rep., {\bf 246}, 315(1994) and
Phys. Lett., {\bf B331}, 137(1994).

\end{references}
\end{document}